\documentclass[fleqn,10pt]{wlscirep}
\usepackage[utf8]{inputenc}
\usepackage[T1]{fontenc}
\usepackage{url}

\usepackage{multirow}

\title{Propagation Dynamics of Rumor vs. Non-rumor across Multiple Social Media Platforms Driven by User Characteristics}

\author[1]{Dongpeng Hou}
\author[1]{Shu Yin}
\author[1,*]{Chao Gao}
\author[1]{Xianghua Li}
\author[1,*]{Zhen Wang}
\affil[1]{Northwestern Polytechnical University, Xi'an City, Shaanxi Province, China}

\affil[*]{corresponding author(s): Chao Gao, Zhen Wang (cgao@nwpu.edu.cn,  w-zhen@nwpu.edu.cn)}

\begin{abstract}

Studying information propagation dynamics in social media can elucidate user behaviors and patterns. However, previous research often focuses on single platforms and fails to differentiate between the nuanced roles of source users and other participants in cascades. To address these limitations, we analyze propagation cascades on Twitter and Weibo combined with a crawled dataset of nearly one million users with authentic attributes. 
Our preliminary findings from multiple platforms robustly indicate that rumors tend to spread more deeply, while non-rumors distribute more broadly.
Interestingly, we discover that the spread of rumors is slower, persists longer, and, in most cases, involves fewer participants than that of non-rumors.
And an undiscovered highlight is that reputable active users, termed `onlookers', inadvertently or unwittingly spread rumors due to their extensive online interactions and the allure of sensational fake news. Conversely, celebrities exhibit caution, mindful of releasing unverified information.
Additionally, we identify cascade features aligning with exponential patterns, highlight the Credibility Erosion Effect (CEE) phenomenon in the propagation process, and discover the different contents and policies between the two platforms. Our findings enhance current understanding and provide a valuable statistical analysis for future research.

\end{abstract}
\begin{document}

\flushbottom
\maketitle

\thispagestyle{empty}


\section*{Introduction}
In the age of digital communication, understanding the dynamics of information propagation in social networks has become increasingly critical~\cite{science2018main,pnas_echo}. It not only sheds light on human behavior and patterns of social interactions but also offers insight into the mechanisms behind the spread of various types of information, from objective news to misinformation~\cite{pnas_behavior,panas_spreading}. This understanding is crucial for a wide variety of applications, including marketing~\cite{science_marketing}, public health communication~\cite{science_Health}, and combating misinformation~\cite{science_Misinformation}. We adopt the traditional definition from the literature on social psychology, which defines a rumor as a story or statement whose truth value is unverified or deliberately false, for better understanding, and others are defined as non-rumors~\cite{ori_rumor,twitterdata2}.

Comprehensive analysis of the characteristic mechanisms of rumors and non-rumors is essential for providing prior knowledge or experience from diverse perspectives.
Some propagation simulation models that quantify or characterize the structural properties and user behavior benefit from the knowledge~\cite{IC_LT}. 
However, some knowledge or experience of analysis often falls short in terms of generalizability due to their typical focus on a single platform~\cite{science2018main, pans_juul2021comparing,science_similar}. 
So the conclusions drawn from such studies may lack persuasive power as they do not account for the variations across different social media sites. 
Moreover, the role of the initiators of information propagation, often termed source users, is significantly different from the participants in the propagation process~\cite{sourcedriven1,sourcedriven2}. However, current works of analysis tend to focus on the characteristics of all users in the cascades indiscriminately, without distinguishing between the attributes of source users and participants~\cite{science2018main, pans_juul2021comparing,science_similar,lerman2010information,PIC1_ICDMprop2013}. This overlooks the potential variations in the behavior and influence of these two distinct user groups, thus limiting the nuance of our understanding of information propagation dynamics from the perspective during the initial (start-up) or spread phase.

In summary, the lack of rigorous multi-platform analysis and the unconsidered facts of the special group of source users (i.e., a purposeful and active disseminator of the message, rather than a passive onlooker) in the cascades of both rumors and non-rumors constitute notable gaps in the existing works~\cite{chen2022information}. 
Our research aims to address these gaps, providing a more comprehensive examination of information propagation patterns across multiple social media platforms.
However, a significant challenge to the study of information propagation in social networks is the limited availability of public data such as user profiles or attributes, due to the privacy restrictions of platforms and other concerns. Key attributes such as the number of followers (fans), number of followings, ratio of followers to followings, number of historical tweets, registration year, and verification status are often missing from publicly accessible datasets~\cite{twitterdata1,twitterdata2,weibodata17}. This lack of data limits the depth of analysis that can be performed and the insights that can be drawn about the role of user characteristics in the propagation of information~\cite{shahi2021exploratory,tu2021rumor2vec}.

To solve these issues, we crawl nearly a million user data records from social media platforms. Using the unique user IDs provided in the propagation data, we match users with propagation nodes, thereby integrating real user information into the public propagation dataset. As a result, we can successfully assemble a comprehensive dataset of information propagation on Twitter and Weibo, which are enriched with authentic user attributes.
This newly integrated dataset allows for a more nuanced and detailed analysis of information propagation patterns, taking into account the role of user characteristics, which facilitates insightful investigations into the dynamics of information dissemination across different social media platforms.
Furthermore, we can dig deeper into this more comprehensive dataset to reveal undiscovered patterns or phenomena based on statistical analysis techniques, such as Complementary Cumulative Distribution Functions (CCDFs),  Maximum Likelihood Estimation (MLE),  Kolmogorov-Smirnov (K-S) test, and chi-squared test. 
A brief summary of some findings is as follows:

\begin{itemize}
\item
Multiple platforms based evaluation of propagation data in Weibo and Twitter is provided in dozens of aspects, including network diameter, propagation depth, propagation breadth, and structural virality~\cite{science2016structural}. 
We conclude that rumors exhibit more viral diffusion, with relatively greater propagation depth, while non-rumors tend to spread in a broadcast-like manner~\cite{science2018main,pans_juul2021comparing}. 
Some different conclusions compared to prior works are that the spread of rumors is slower, persists longer.
And though we analyze that the average number of participants in rumors exceeds that in non-rumors, aligning with people's common perception, it is noteworthy that 90\% of mundane or trivial rumors actually involve fewer participants compared to non-rumors.


\item
An analysis of user attributes across various dimensions in the distinct view of the source group and participant group has been implemented. The results show that the quality of users who initiate non-rumors is generally higher than those who start rumors.
However, the conclusion drawn from the group of participants is the opposite.
In more specific terms, we find that source users who initiate non-rumor information tend to be individuals with considerable influence, such as those with a large number of fans. These individuals, often public figures, may be reluctant to spread rumors due to potential negative consequences. 
On the contrary, participating users who propagate rumors may have higher levels of activity on social media. We can conclude that high-profile celebrities may be careful about sharing inaccurate information, and attractive rumors often attract user groups with higher activity levels on social media.

\item
Some unique social phenomenon findings are unearthed. (1) The propagation features conform more closely to an exponential distribution. (2) The propagation process exhibits a phenomenon known as the Credibility Erosion Effect (CEE). This means that the credibility of an individual, who frequently broadcasts and shares the same information, gradually declines over time. (3) The influence of different content topics on propagation across different platforms varies. For instance, on Weibo, rumors are often news-related, while non-rumors tend to be entertainment-related. However, among non-rumors, the content that spreads quickly and deeply is often sensational news.
Contrary to Weibo, both rumors and non-rumors involve an amount of news on Twitter. 
And different from Weibo, where non-rumors' propagation depth far surpasses rumors, we do not observe such a phenomenon on Twitter. These discrepancies also reflect the different mechanisms and policies for rumor detection across various platforms.

\end{itemize}

\section*{Results}
In our analysis, each cascade within social networks is treated as an independent propagation graph, then the network science and statistical  techniques can be applied to these real-world cascades. 
This strategy allows us to conduct a comprehensive evaluation from three distinct perspectives: the topological structure mapped from the propagation graph itself, the directed propagation graph originating from the source, and the various attribute indicators of users on each independent propagation.
To succinctly encapsulate these complex dynamics, we employ CCDFs to conduct the statistical analysis rigorously. CCDFs offer several advantages for the analysis of such data. Firstly, they are adept at handling large datasets and large variations in data values. Secondly, they can intuitively represent the frequency of occurrence of a wide range of values in the dataset. Therefore, employing CCDFs, we can visually capture and compare the intricate patterns and trends in both rumor and non-rumor cascades.

\subsection*{Statistics based on the Topology of Propagation Cascades: Broadcast Spread based Non-Rumors and Viral Spread based Rumors}
\begin{figure}[ht]
\centering
\includegraphics[width=\linewidth]{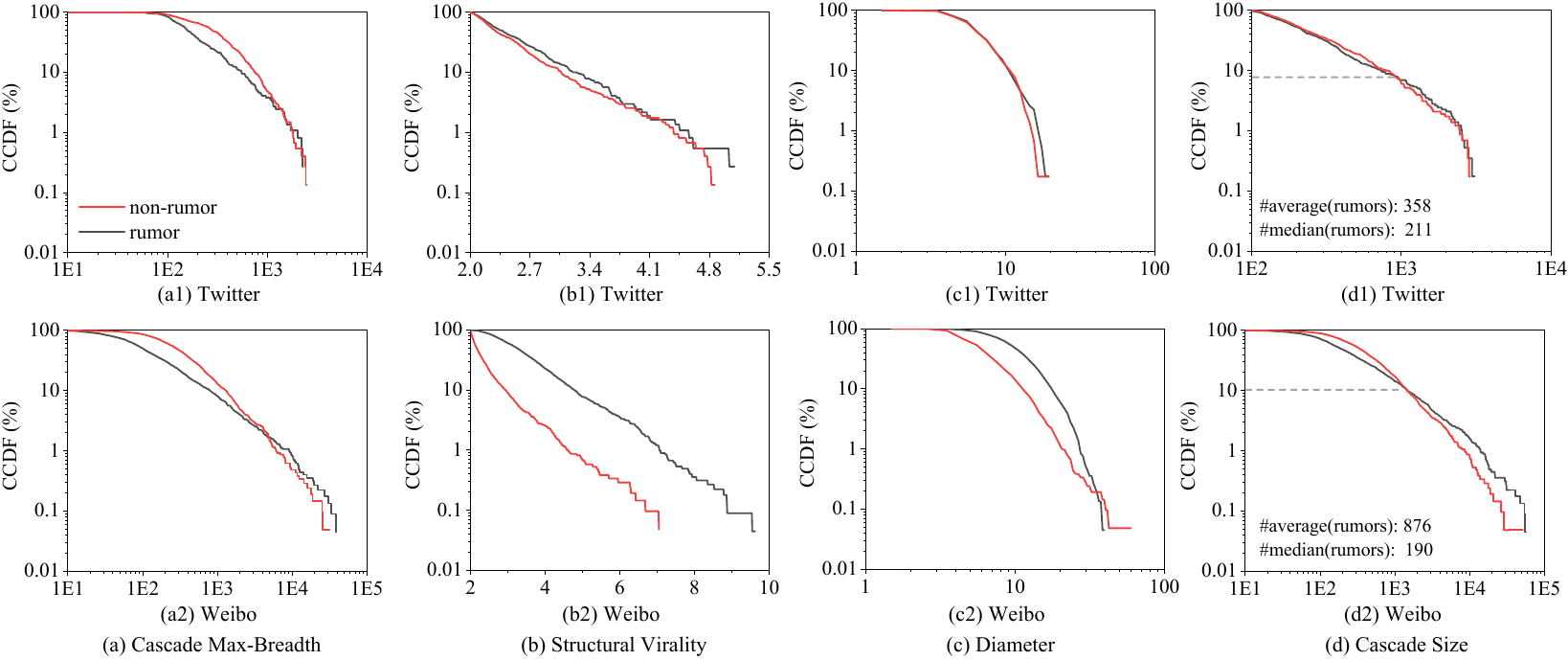}
\caption{Complementary cumulative distribution functions (CCDFs) statistics of critical network topology-based attributes in rumor and non-rumor cascades for Twitter (the first row) and Weibo (the second row).
From left to right, the plots represent (a)  the maximum breadth, (b) the structural virality score, (c) the diameter, and (d) the cascade size. 
In all cases, the CCDFs related to broadcasting (Max-Breadth) for non-rumors are higher than that for rumors. In contrast, rumors are beyond non-rumors when associated with viral diffusion (Structural Virality and Diameter).
About 90\% of rumors involve fewer participants compared to non-rumors.
However, the average number of participants in rumors is larger than that in non-rumors.
}
\label{fig1}
\end{figure}

First, Fig.~\ref{fig1} presents the CCDFs of network topology-based rumor and non-rumor cascades for Twitter and Weibo. 
These CCDFs provide a concise representation of the patterns of critical structural attributes related to both rumor and non-rumor events across these two platforms.
Our analysis concentrates on four specific aspects of cascade topology: max-breadth, structural virality, depth, and cascade size~\cite{science2018main}, which represent the crucial topological features of information propagation.
The max-breadth refers to the maximum degree among all nodes in the cascade at any depth level, representing the widest point of the cascade. This shows the highest number of users simultaneously engaging with or spreading the information, offering insight into the peak intensity of the information spread. Structural virality measures the complexity of the propagation graph, representing how branched the dissemination process is. 
Diameter refers to the longest shortest path between any two nodes in the graph, representing the maximum distance across the underlying topology of the network. 
Lastly, the cascade size represents the total number of participants in a given propagation, reflecting the general extent of the spread of the information.

As can be seen in Fig.~\ref{fig1}, the trends of CCDFs reveal some interesting patterns. In the case of max-breadth, non-rumor cascades tend to rise above rumor cascades, suggesting a broader initial spread for truths than for falsehoods. Conversely, in the structural virality and depth graphs, rumor cascades predominantly sit above non-rumor cascades, indicating a higher complexity and depth in the propagation of rumors. 
Lastly, for the cascade size (i.e., number of participants), a different phenomenon is revealed.
the CCDF trend of non-rumor cascade size overall exceeds rumor cascade size, which is opposite to people's daily cognition that the influence of rumor is relatively large. Therefore, we also compute the average number of participants in rumor and non-rumor cascades. The conclusion is that the average level of rumors exceeds that of non-rumors. For example, in the Weibo dataset, the average number of participants in rumor cascades is \textbf{876}. In comparison, the average number of non-rumor cascades is \textbf{710}.
By further exploring the intersection point of two CCDF trends in Fig. 1 (d), we conclude that:

\begin{enumerate}
    \item 
Trivial rumors, which constitute 90\% of all rumors, have fewer participants than non-rumors. In contrast, the remaining 10\% of rumors are sensational in nature, attracting a significantly larger number of participants. 
This small proportion of highly engaging rumors substantially raises the average level of participants in rumors, even resulting in the average number of participants for all rumors being greater than that for non-rumors.
From the above perspective, the average number of participants in rumors is markedly higher than the median (The post hoc analysis reveals that the medians of rumor participants in the Twitter and Weibo platforms are only 211 and 190, respectively).
 \item 
The discrepancy between public perception and actual data can be attributed to media influence. The media often focuses on and reports sensational rumors due to their eye-catching content, which, despite their rarity, leave a lasting impression on the public. In contrast, other mundane rumors receive less attention.
 This selective and one-sided reporting leads to an overestimation of the overall impact of rumors. In reality, when these sensational rumors are excluded, the average participation in the remaining 90\% of rumors is actually lower than that in non-rumor events. 
 \item 
In summary, the average number of participants in rumors is greater than the median, indicating that the influence and reach of most rumors are less extensive than commonly believed. Moreover, a conservative reason might be the intensified efforts on various platforms to identify and counter rumors~\cite{counter_debunk1,counter_debunk2,counter_debunk3,counter_debunk4,counter_debunk5}, which leads to many rumors being debunked or restricted early in their spread, thereby preventing them from achieving large-scale spreading\footnote{More evidence is analyzed in the third part of the Discussion section.}.

\end{enumerate}

\begin{figure}[ht]
\centering
\includegraphics[width=0.6\textwidth]{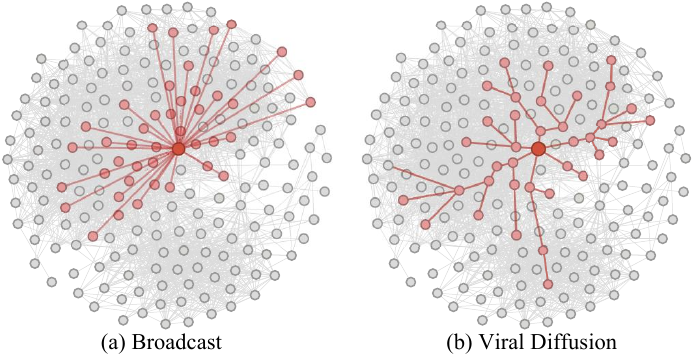}
\caption{Comparison between broadcast and viral diffusion. While broadcast diffusion typically involves a central source that spreads information uniformly to a wide audience, viral diffusion is characterized by organic, peer-to-peer spread, and information cascades often show larger depth and diameter.
}
\label{fig_gephi}
\end{figure}

Based on our analyses, we can draw several key conclusions about the propagation dynamics of rumors and non-rumors in social media. When it comes to non-rumor cascades, their larger max-breadth suggests a broader immediate impact, which is characteristic of a broadcast-like spread. This indicates that verifiable information tends to quickly reach a wide audience, facilitated by the network's interconnected structure. The illustration of broadcast based non-rumors is presented in Fig.~\ref{fig_gephi}(a).
On the other hand, rumor cascades, characterized by greater depth and structural virality, demonstrate a more complex and far-reaching spread. This implies that the propagation of rumors often undergoes intricate branching paths, reaching deeper into the network over time. The higher structural virality of rumor cascades suggests a more viral-like spread, where rumors can permeate through the network via multiple routes, lending it a persistent and pervasive presence.  The illustration of viral diffusion based non-rumors is presented in Fig.~\ref{fig_gephi}(b).
Therefore, while non-rumors tend to quickly reach a broad audience in a more straightforward manner akin to broadcasting, rumors often spread deeper and in a more complex way, much like a virus infiltrating a host organism.

\subsection*{Statistics of the Directed Graphs from Sources: Slower and More Persistent Nature of Rumor Propagation}
\begin{figure}[ht]
\centering
\includegraphics[width=\linewidth]{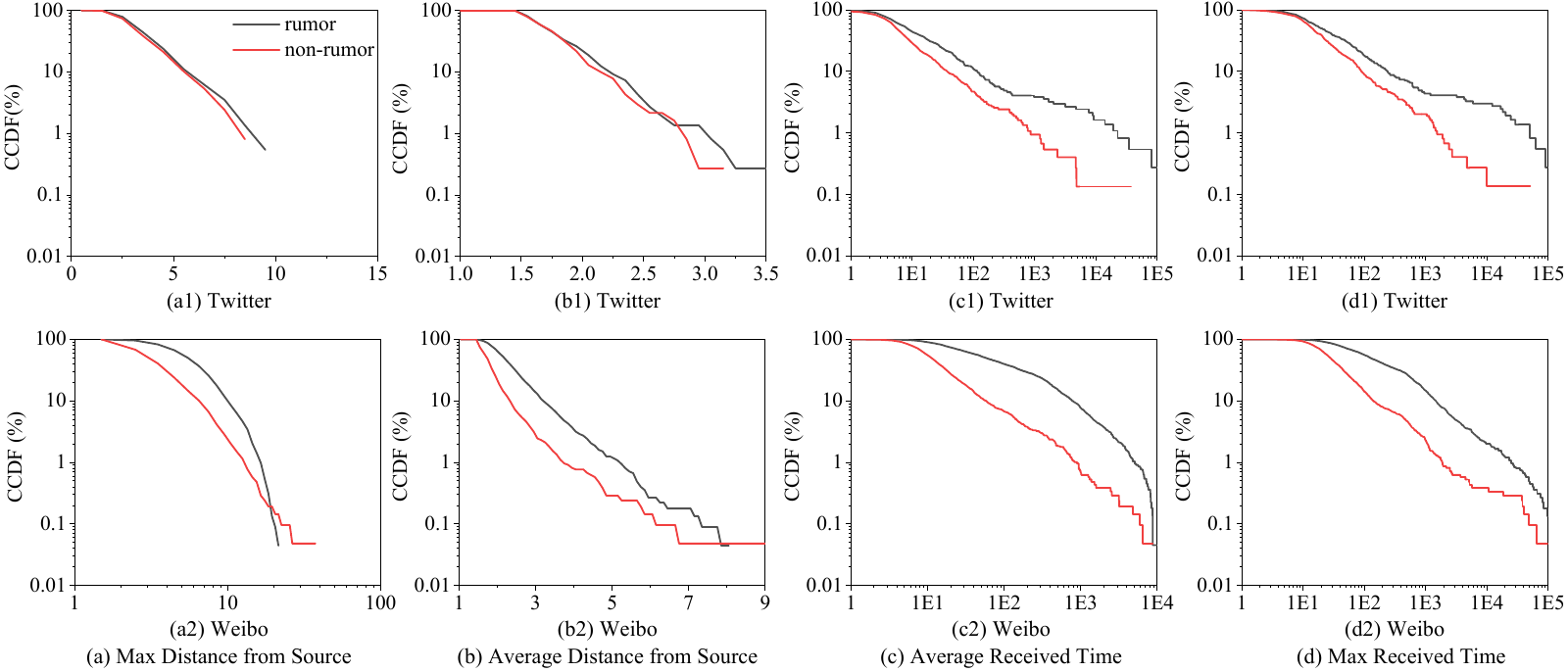}
\caption{CCDFs of four propagation attributes for rumor and non-rumor cascades on Twitter (the first row) and Weibo (the second row). From left to right, the plots represent (a)  the maximum hop distances, and (b) the average hop distances from the source user. (c) the maximum time, and (d) the average time taken for the cascades to reach other users after being spread from the source. In all cases, the CCDFs for rumor cascades are plotted above those of non-rumor cascades, indicating deeper, slower but persisting propagation patterns for rumors. }
\label{fig2}
\end{figure}

In this part, we transition from analyzing static topological attributes to scrutinizing the dynamic process of information cascades, specifically considering the propagation of information from its source outwards.
There are four distinct propagation dynamic attributes for each platform, which include the maximum and average hop distances from the source, and the maximum and average time taken for the cascades to reach other users after being disseminated from the source.
Here, we demonstrate a series of four distinct attributes. The attributes we consider are the maximum distance from the source, the average distance from the source, the maximum time taken for other users to receive the information after the information has been disseminated from the source, and the average time taken for the information to spread across the entire cascade.
As depicted in Fig.~\ref{fig2}, all eight graphs present the CCDFs of these attributes for rumor and non-rumor cascades in two platforms. 
In each case, the CCDFs of rumor cascades are consistently above those of non-rumor cascades.

More specifically, the maximum and average hop distances from the source serving as depth indicators refer to the longest path from the root of the cascade to any leaf, indicating the farthest reach of the information spread. A greater distance suggests that the information, whether rumor or non-rumor, has penetrated further into the cascade. Interestingly, the larger depths for rumor cascades in these distance attributes from the source echo the previous findings of greater diameters in the topological analysis, indicating that rumors tend to disseminate deeper into the cascade.
On the temporal front, the maximum and average times taken for the cascades to reach other users measure the speed of the propagation. Longer times could indicate a slower, yet potentially more continuous spread, this characteristic is often associated with rumors. Such a finding corroborates our prior observation of higher structural virality in rumor cascades, suggesting that the time taken for information to spread is intimately tied to the complexity of its propagation path. Moreover, the slower spread of rumors, as represented by longer reception times, resonates with the smaller cascade size observed previously, indicating fewer recipients over the same time frame.
In summary, the propagation dynamics observed from the source are consistent with our earlier findings from the topological analysis. Rumors, with their deeper penetration and slower spread, reveal a persistent\footnote{
We also rigorously use an indicator linking each depth from the source to the average information receipt time at that depth~\cite{science2018main} to measure the average and maximum time for rumors and non-rumors at each depth in the propagation chain. The corresponding content is in the third part of the Discussion section.} and pervasive presence in the cascade. This underscores the inherent difficulties in curtailing the spread of rumors, given their tendency to infiltrate deeper and persist longer in the cascades.

Connecting the insights from both the topological and dynamic analyses, we can conclude that rumor and non-rumor cascades exhibit distinct behaviors on social media platforms. Rumors tend to spread deeper and slower, reaching fewer individuals over a given time but persisting in the cascade longer. Non-rumors, on the other hand, spread faster and wider, reaching a larger audience quickly but not penetrating as deep. These findings offer an intuitive understanding of the propagation patterns of rumors and non-rumors in social media, which is critical for designing effective strategies to mitigate the spread of rumors. 
What's more, not only do we thoroughly analyze the physical patterns and trends implied in rumors and non-rumors propagation cascades from the perspective of real-world characteristics, but we also find the convexity pattern of most of the trends in the CCDFs and the corresponding concavity pattern in Cumulative Distribution Functions (CDFs) from Fig.~\ref{fig1} and Fig.~\ref{fig2}.
This prompts us to analyze the distribution (such as exponential, gamma, or log-normal pattern) from the perspective of numerical analysis, which is demonstrated in the first part of the Discussion section.

\subsection*{Statistics based on the User Attributes: Cautious Influential Source Users and Active Onlookers in Rumor Propagation}
Having explored the network's structural attributes and the dynamic process of information propagation, we now turn our attention to the individual users within these information cascades. Each node in the cascade represents a user, identified by a unique user ID (UID). Given the importance of privacy, existing propagation data do not disclose individual attributes. Therefore, we collect comprehensive data from nearly a million users participating in these cascades. This data set includes several critical attributes, such as the number of fans, the number of followings,
the number of historical tweets, the year of registration and the verification status.
By mapping each node in the cascade with these real-world user attributes, we can reconstruct a more accurate and complete representation of the cascade. This strategy allows us to analyze the spread of information in relation to the users' social influence, activity level, and credibility within a network, and provides a comprehensive understanding of how user characteristics may influence the propagation of rumors and non-rumors on social media platforms. 

Although existing research has analyzed user attributes, these studies often treat the source users and participants as the same group, providing a unified analysis of all user roles in information propagation. Therefore, these works overlook a vital aspect of information propagation: the unique characteristics of the source. However, (1) from a sociological perspective, the initiators of the cascade, the sources, are inherently directional. Then, they spread information driven by specific motivations or considerations.
Therefore,  understanding the attributes of sources can help in predicting the potential spread of a piece of information, its trajectory through the network, and the nature of the cascade it could generate.
For example, a source with a large number of followers may initiate a cascade that spreads far and wide. Similarly, the credibility of the source, as indicated by its verification status, could influence the rate at which its information is accepted and propagated by others.
(2) Delving into the characteristics, motivations, and behaviors of these sources can furnish valuable insights into the initiation of information cascades. Particularly involving rumors, recognizing these source attributes can play a pivotal role in early rumor detection and mitigation. 
(3)  Furthermore, the source user group and the participant group respectively illuminate the intentions underpinning the initiation and subsequent overall spread of rumors or non-rumors. 
Understanding the varied phases of information propagation provides crucial insights for effective prevention and interruption (from the source's perspective during the initial or start-up stage), and wide unwitting public awareness campaigns (from the participant's perspective during the spreading stage).
This distinction between the source and the participants is the highlight of our study, which allows us to shed light on an underexplored yet critical facet of information propagation, enriching our understanding of how rumors and non-rumors spread in social media networks.
Mainly, for our study, we consider a number of cascades. Each cascade, affiliated with either rumors or non-rumors, comprises a collection of users with UIDs and the corresponding attributes. We examine attribute statistics for the source users and the average level of participants within rumor and non-rumor cascades.
In the following, we detail the findings in two distinct groups.


\begin{table}[ht]
\centering
\caption{Comparison of user verification status in propagations related to true and false events (non-rumor and rumor) for Twitter and Weibo. 
The source group only counts the verification status of the source users for each propagation, while the participant group includes the verification status of all users involved in each propagation.}
\label{tab1}
\setlength{\tabcolsep}{3.5mm}{
\begin{tabular}{|c|cc|cc|}
\hline
\multirow{2}{*}{\centering Condition} & \multicolumn{2}{c|}{Source Group} & \multicolumn{2}{c|}{Participant Group} \\
\cline{2-5}
 & Rumor & Non-rumor & Rumor & Non-rumor \\
\hline
Twitter & 0.7270 & 0.8914 & 0.2182 & 0.2139 \\
\hline
Weibo & 0.3685 & 0.8775 & 0.0430 & 0.0481 \\
\hline
\end{tabular}
}
\end{table}

We consider the verification status (authority) of users involved in the propagation of rumors and non-rumors. If we follow existing methodologies and only calculate the verification ratio of all users participating in either rumor or non-rumor cascades, we find that, within each platform, the verification ratios for rumors and non-rumors are comparable. However, when we additionally examine the source group, we observe a striking and somewhat expected difference: the verification ratio of users who initiate the spread of non-rumors is significantly higher than that of those who spread rumors. This difference is particularly pronounced on the Weibo platform, where the verification ratio of sources spreading non-rumors is 138\% higher than that of sources spreading rumors.
Tab.~\ref{tab1} provides a detailed comparison of the status of user verification in the propagation of true and false events for both Twitter and Weibo. Here, the source group only counts the verification status of the source users for each cascade, while the participant group includes the verification status of all users involved in each cascade.
Notably, our approach of considering the source separately aligns with the sociological perspective that the initiators of information propagation are not just passive transmitters but active agents driven by certain motivations. 
By focusing on the source, we can glean unique insights that might be obscured when analyzing all participating users as a homogenous group. 
We can conclude that the distribution in terms of verification characteristics related to rumor and non-rumor propagation in the source groups is opposite to that of the participant groups. 
Furthermore, the significantly higher verification ratio of source groups from non-rumor cascades suggests that verified users, who are often more credible and influential on social networks, are less likely to initiate the spread of rumors. This finding also implies that the verification status of the source could serve as a valuable feature for early detection and intervention of rumors.

\begin{figure}[ht]
\centering
\includegraphics[width=\linewidth]{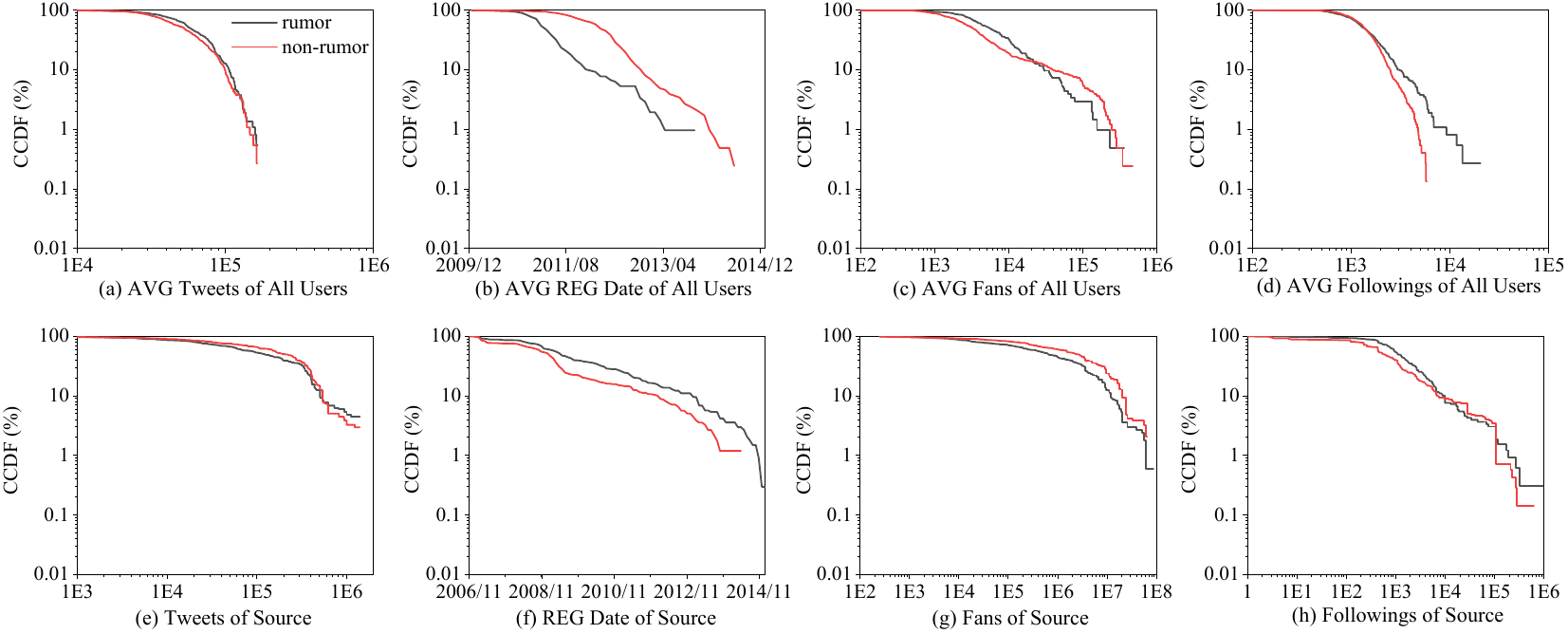}
\caption{CCDFs of four user attributes for rumor and non-rumor cascades on Twitter. From left to right the plots represent (a) the number of tweets, (b) the registration date, (c) the number of fans, and (d) the number of followings.
In all cases, the comparison trend between rumors and non-rumors is largely inverse when observed at the average level of all participants (the first row) and at the source user level (the second row).
Only the metric of followings shows an insensitive trend between rumor and non-rumor for both the source user group and the participant group.
}
\label{fig3}
\end{figure}

\begin{figure}[ht]
\centering
\includegraphics[width=\linewidth]{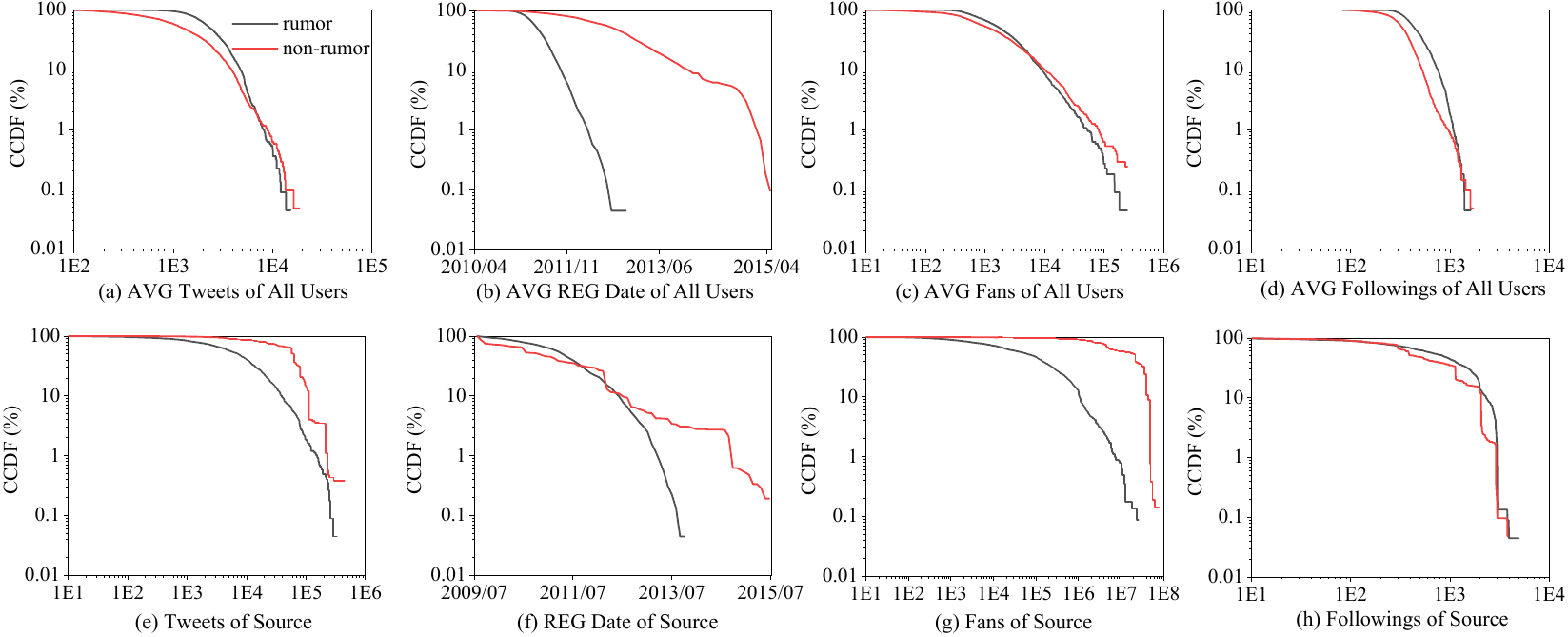}
\caption{CCDFs of four user attributes for rumor and non-rumor cascades on Weibo. Similar to Fig.~\ref{fig3}, the comparison trend between rumors and non-rumors is largely inverse when observed at the average level of all participants (the first row) and at the source user level (the second row). And only the metric of followings shows an insensitive trend between rumor and non-rumor for both the source user group and the participant group.
}
\label{fig4}
\end{figure}

Similarly, we further analyze the source user group and the participant group across four additional user attributes, focusing on their corresponding cascade's CCDFs. To highlight the differences between source and participant groups, we present the results for the two platforms separately in two figures. Each figure illustrates the comparison between the source user group and the participant group across the four dimensions of a platform.
\begin{itemize}
    \item  \textbf{Number of Tweets (Tweets Activity)}: It can be seen from Fig.~\ref{fig4}(a) that within the participant group, a higher tweet count is associated with users involved in the propagation of rumor cascades. This potentially implies that the spread of rumors leverages the extensive, grassroots masses in social media, emphasizing the role of the general public in the diffusion of rumors. Intriguingly, when we turn our attention to the source users in Fig.~\ref{fig4}(e), those who initiate non-rumor cascades conversely exhibit an intensified tweet activity. This observation might imply that the propagation of non-rumor based content is primarily launched and started by a minority group of more active users. These small numbers of users are often recognized and influential (celebrity) within their social platforms, likely due to their cautious approach toward mitigating the detrimental outcomes of spreading rumors and their consistent contribution of high-value content.

    \item 
\textbf{Registration Date (User Seniority)}: According to Fig.~\ref{fig4}(b), participants in rumor cascades are more likely to have a longer registration duration on the platform,  indicating a tendency for rumors to be spread by more established users. Turning attention to the registration date of the source group in Fig.~\ref{fig4}(f), the source users of non-rumor cascades are observed to be even longer-standing members of the platform, presenting a potential credibility or trust associated with account longevity that tends to initiate the propagation of non-rumor content, while newer users with less experience are more likely to become the rumor sources, due to their inadequate familiarity with the platform's norms and fact-checking resources, or their greater susceptibility to fall for incorrect information

   \item 
\textbf{Number of Fans (Popularity)}: According to Fig.~\ref{fig4}(c), the larger fanbase among participants in rumor cascades demonstrates that rumors may spread more easily among users with a wider audience reach. In contrast in Fig.~\ref{fig4}(g), source users in non-rumor cascades have a larger number of fans, possibly indicating their established status or perceived credibility within the social platform.
This may indicate that high popularity users are more careful when sharing unverified information due to their visibility and reputation risk.

   \item 
\textbf{Number of Followings (Information Seeking)}: Referring to Figs.~\ref{fig4}(d) and (h), a consistent trend is observed across both the participant and source groups, where users in rumor cascades have a higher following count. This demonstrates that more comprehensive and extensive information receiving behaviors enable users to widely receive both rumor and non-rumor contents. However, due to unique characteristics such as sensationalism or novelty, rumors may exhibit greater attractiveness. As a result, higher following users, whether they are as sources or participants, tend to select to be involved in attractive rumor-related events. This aligns with the actual psychological propensity towards sensational or novel information. Then we can conclude that more comprehensive and extensive information receiving behaviors could potentially increase the likelihood of users encountering and participating in the spread of rumors.

\end{itemize}



In summary,  the distribution of most of the user features related to rumor and non-rumor propagation in the source groups is opposite to that of the participant groups. 
Among them, the base of rumor propagation lies in the users with active profiles with decent credibility. These users, often referred to as the grassroots or onlookers, exhibit good reputations and generally active metrics across various aspects (such as a larger number of tweets, and higher followings to Internet surfing\footnote{The dominant weight of different characteristics in collected user profiles on the propagation of rumors or non-rumors is demonstrated in the second part of the Discussion section.}). However, their conviction in the veracity of information lacks the level of certainty that the initiators possess. As a result, these broad user groups are more susceptible to the allure of attractive rumors. Their participation lends momentum to the propagation of rumors, highlighting the necessity for robust fact-checking mechanisms and awareness programs to enhance supervision among such grassroots based users with appropriate credibility, aiming to mitigate the spread of rumors.
On the other hand, the behavior of the source user group, who initiates the information, presents a contrary trend. Specifically, the sources in non-rumor cascades often exhibit higher credibility (earlier registration) or influence (larger number of fans) than the sources in rumor cascades. This distinction suggests that users, particularly those who have a large number of fans, are circumspect about spreading rumors when they possess a greater degree of certainty or awareness of the information's veracity. 
These high-quality users, driven by a desire to maintain their reputation, promote positivity, or uphold ethical standards, demonstrate a stronger inclination toward sharing truthful information. Such a tendency highlights its pivotal role in preventing the genesis of rumors.

\section*{Discussion}
\subsection*{Exponential Distribution Phenomena of Cascade's Topological Characteristics }
From Fig.\ref{fig1} and Fig.\ref{fig2}, which illustrate the CCDFs of topological characteristics of rumor and non-rumor cascades, we can infer the trend of the CDF.
The visibility of the convex function trend of CCDFs or the concave function trend of CDFs of these features is also evident. 
The CDF of topological characteristics, including breadth, depth, and structural virality, among others, usually appears as a smooth curve above the $y$=$x$ line. This pattern suggests that the data is possibly sourced from a distribution that is positively skewed or exhibits a long tail.

After attempting to fit the cascade data's topological features into several distributions, we find that the exponential distribution is the one that best matches our dataset\footnote{A detailed description of the methodology approach is provided in the second part of the Methods section.}. Further rigorous testing is then conducted to validate the appropriateness of this fit. Characteristics like structural virality produce p-values exceeding a 0.05 significance level. Hence, for these particular cascade characteristics, the exponential distribution is a suitable fit.

\subsection*{Ranking of User Feature Importance in Rumor vs Non-Rumor Cascades}
To discern which categories of user features are the most impactful in determining whether a cascade is rumor associated or not, we perform a chi-squared test analysis to provide a quantitative measure of the importance of each feature~\cite{chi-squared}, thus elucidating the key factors influencing whether a cascade propagates rumors.
Our feature set statistically computes five metrics for each cascade: the average number of tweets, the average registration date, the average number of followings, the average number of fans, and the proportion of verified users participating in the cascade. 
By sorting the features according to their chi-squared statistics in descending order\footnote{A detailed description of the methodology approach is provided in the third part of the Methods section.}, we obtain a ranking of feature importance: 
\textbf{number of fans}, 
\textbf{verification status}, 
\textbf{registration date}, 
\textbf{number of tweets},
\textbf{number of followings}.

\begin{enumerate}

\item 
Similar to the celebrity effect and in line with people's intuition, the number of fans indeed proves to be the most significant indicator.
The top ranking of fans number could be attributed to the larger social influence of the users having more fans. These users are more likely to be trusted by their followers, so if such a user propagates a rumor, it can spread quickly and widely. In the same way, a large influence can also limit their behavior due to the influence of negative public opinion.

\item 
The second ranking of the user verification status indicates that verified users are less likely to propagate rumors, perhaps due to their higher public visibility and greater potential damage to their reputation.
Empirically, we consistently recognize that this indicator plays an important role in the propagation of rumors in social media.
However, its moderate ranking instead of top ranking also reveals an interesting aspect of rumor cascades, which aligns with our earlier analysis. 
As evidenced in Tab.~\ref{tab1}, the verification status of the initiating user, the source of the cascade, plays a crucial role in determining whether the cascade disseminates rumor based or non-rumor based content. Because the source users, as the drivers of cascades, often possess a deeper understanding of the rumors in question.
In contrast, when we consider all users involved in the cascade, the impact of the verification status becomes less pronounced. This can be attributed to the behavior of participating users, who often act as innocent onlookers in these scenarios. Driven more by the allure of participating in trending topics and engaging narratives, these users may not exercise the same level of scrutiny as the source users in regard to the veracity of the information they propagate.
Hence, when we examine the entirety of a cascade, the individual impact of verification status becomes diluted, resulting in its moderate importance ranking.

\item 
The third ranking of registration date suggests that when a user joins the corresponding platforms is a strong indicator of whether they will participate in rumor propagation. One possible reason could be that accounts with early registration have understood the platform's norms and the bans for spreading rumors, making them less likely to propagate rumors.

\item 
The number of tweets and average number of followings ranking last suggests that these factors are less indicative of rumor propagation.
Nevertheless,  users who tweet more frequently could have a higher chance of spreading rumors simply due to their higher volume of output. However, these factors are less decisive compared to the others.  
It is worth mentioning that users who follow a large number of other accounts could potentially be exposed to a wider range of information, including rumors. 
However, the reason for the insignificance of this indicator as the last in ranking aligns with the consistent conclusions drawn from the evidenced Figs.~\ref{fig3}(d), ~\ref{fig3}(h), ~\ref{fig4}(d), and ~\ref{fig4}(h), where the propensity towards rumor propagation does not show a contrary trend between the source group and the participant group.

\end{enumerate}

These findings provide valuable insights into the behaviors of users involved in rumor propagation, which could be leveraged to develop more effective strategies for rumor detection and prevention on social media platforms.

\subsection*{Disparate Causes of Identical Rumor Spreading Patterns on Twitter and Weibo}
\begin{figure}[ht]
\centering
\includegraphics[width=0.92\linewidth]{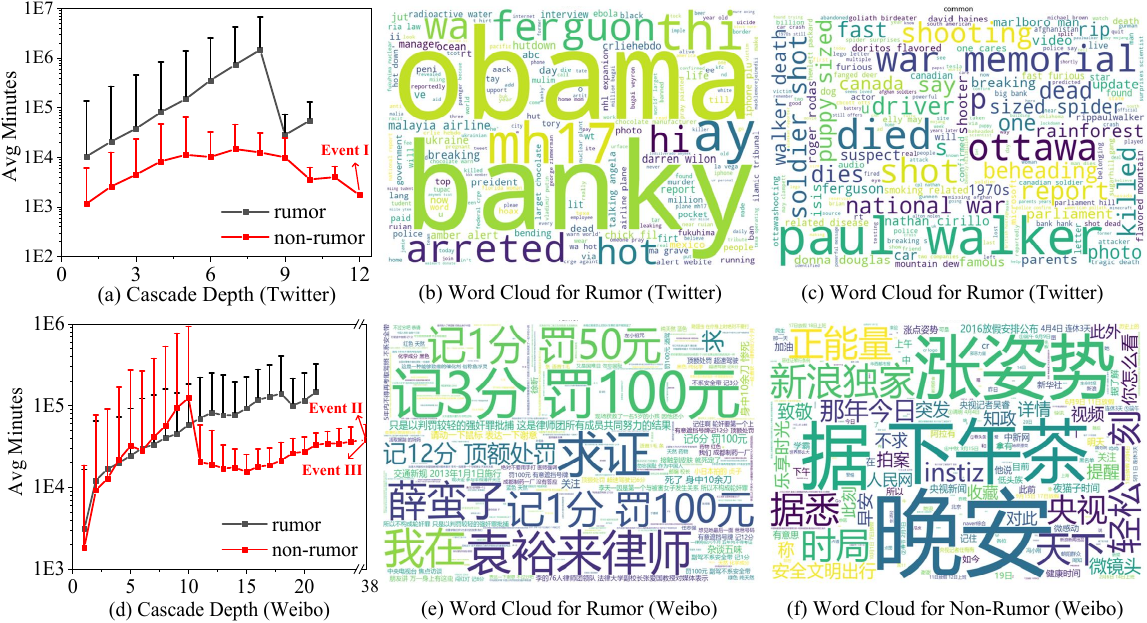}
\caption{Spreading patterns analysis (propagation speed comparisons and topic content analysis) for rumors and non-rumors on Twitter (the first row) and Weibo (the second row). From left to right: (1) the first column showcases the speed of propagation for rumors vs. non-rumors, (2) the second column presents the word cloud for rumors, and (3) the third column depicts the word cloud for non-rumors. 
}
\label{fig5}
\end{figure}

Despite presenting similar patterns of rumor propagation in Fig.~\ref{fig2} and Fig.~\ref{fig3} (such as slower,  deeper, persistent, and high structural virality), Twitter and Weibo display disparate underlying reasons. We examine an essential metric that quantifies the average time taken for a rumor or non-rumor to propagate from the source to every subsequent layer of depth, as depicted in Figs.~\ref{fig5} (a) and (d). 
In Fig.~\ref{fig5}(a), the propagation of non-rumor in each depth is overall faster than that of rumor on Twitter, indicating a slower yet more persistent spread for the rumor propagation, which has been verified consistently in Figs.~\ref{fig2} (c1) and (d1). 
In contrast, Fig.~\ref{fig5}(d) presents a different divergence from Figs.~\ref{fig2}(c2) and (d2) in the early propagation stage, that is, rumors and non-rumors exhibit similar propagation patterns instead of non-rumor faster than rumor.
Not only that, we also find that the maximum propagation depth of the two platforms is also very different.

To delve deeper into the differences, we rigorously performed a word cloud analysis on the content, as shown in the middle and right columns of Fig.~\ref{fig5}. 
Notably, the topic content on Twitter and Weibo varied considerably. The content released from Twitter, whether rumors or non-rumors, is mostly related to politics, social crime news, etc. 
However, Weibo displays distinct topics between rumors and non-rumors. 
Rumors are mostly about social issues, people's livelihoods, crime news, etc., while non-rumors are less about significant news and more about entertainment and light-hearted topics.
Then the reasons for some special phenomena can be discovered.

\begin{itemize}
\item 
Maximum Propagation Depth of Non-rumor but with Smaller Structural Virality: The maximum propagation depth is higher for non-rumors, which typically correspond to significant international or domestic events. Specifically, as red markers highlighted in Fig.~\ref{fig5}, examples of these events include the American presidential election opinion poll (Event I), the tragic incident in which a Heilongjiang police officer was fatally shot by a criminal (Event II), and the award ceremony for Nobel Prize laureate Tu Youyou (Event III).
However, these events are relatively few and insufficient to significantly drive the average level of structural virality. 
Therefore, rumors tend to have a higher structural virality at an average level, while non-rumors typically possess a very handful of highly impactful events with larger depth and faster propagation. Although the structural virality of these non-rumor events is more pronounced, their quantity is very limited.

\item 
Differences in Maximum Depth: The differences are primarily manifested in two aspects. First, the maximum depth for both rumors and non-rumors on Twitter is similar, showing no significant disparity. In contrast, on Weibo, the maximum depth of non-rumors is significantly larger than that of rumors. Secondly, the overall maximum depth on Twitter is considerably lower than on Weibo. 
These disparities could be attributed to differences in thematic content on the platforms, leading to variations in their respective detection mechanisms.
More specifically, rumors on Weibo predominantly revolve around news topics, as opposed to the usual entertainment content of non-rumors, and may involve political factors. As such, these sensitive rumors are timely curbed after the intervention of the official detection mechanism, resulting in a significantly shallower propagation depth for rumors on Weibo compared to non-rumors.
On the contrary, Twitter does not exhibit instances of extremely deep cascades. This could be partially attributed to the high similarity in topics between rumors and non-rumors on the Twitter platform.

    \item 
Differences in Early Propagation Speeds: 
The topic dimension on Twitter of rumors and non-rumors are all similarly related to news, as evidenced by Figs.~\ref{fig5}(b) and (c), does not have a divergent impact on the spread of information on this platform, so the trends in Figs.~\ref{fig2} (c1)(d1) and Fig.~\ref{fig5}(a) are consistent. That is, non-rumors spread faster than rumors on average.
In comparison,  as evidenced by Figs.~\ref{fig5}(e) and (f), the rumors on Weibo have a news tilt and the non-rumors usually are entertainment.
Rumors that lean towards news topics are more sensational, therefore, they spread faster in the early propagation stage (Fig.~\ref{fig5}(d), 1-10 depths).
This difference in topics has accelerated the early-stage propagation speed of rumors on Weibo compared to Twitter, resulting in the downward shift of the black line in Fig.~\ref{fig5}(d) relative to Fig.~\ref{fig5}(a).
However, the observed overall faster non-rumor propagation in Figs.~\ref{fig2} (c2) and (d2) is due to a few deeply impactful true events, which are characterized by high hot-topic and result in deeper and faster propagation (such as event II and III). 
These hot-topic based non-rumors with fast spreading can push and reduce the average and maximum propagation time of the overall non-rumors on Weibo.

\item 
Inflection Points Phenomenons in Non-rumor Cascades: Both platforms have inflection point phenomena for nonrumor events. This occurrence is due to highly popular non-rumor events that spread incredibly quickly. The average propagation speed for the small depth (hops) layer from sources appears slower due to the influence of a larger number of non-rumor events with a maximum shallower depth. Once the depth reaches the inflection point, the propagation time drops significantly, indicating influential events. The occurrence of these rare and unique events is rare, yet they have a significant influence on the propagation dynamics.

\end{itemize}

So we can conclude that the differences in thematic content across the two platforms and the different mechanisms for rumor detection of the two platforms cause noticeable disparities in cascade depth.
(1) The news-driven rumors on Weibo compared with entertainment content for non-rumors attract more participants and cause the faster spreading of rumors in the early stage. 
There is no doubt that their sensationalism is also more closely monitored by the detection mechanisms of the platform. 
Therefore, these rumors are often quashed by the platform before they can cause a significant impact (e.g., achieve substantial propagation depth). Furthermore, the maximum depth of rumors on Weibo is much less compared to non-rumors.
(2) The reason non-rumor events on Weibo can propagate more deeply and quickly is that these unique and very few long-depth-based non-rumor events are also oriented toward news (the red highlighted events II and III in Fig.~\ref{fig5} have been evidenced).
In contrast, Twitter, overall having news-biased content, shows a general trend of faster real news propagation and slower fake news spread. Overall, the propagation process aligns with human intuition that news is more time-sensitive.
(3) The platform's stringent measures against rumors make it challenging to collect data on rumors with large-scale participation. 
Potentially, this could also explain the discrepancy between our statistical results (where rumors only surpass non-rumors in terms of average participant numbers and 90\% of rumors have fewer participants compared to non-rumors as depicted in Fig.~\ref{fig1}(d)) and common perception (which assumes that participation in rumors should overwhelmingly exceed that in non-rumors).

\subsection*{Credibility Erosion Effect Phenomenon in Propagation of Social Media}
The Credibility Erosion Effect (CEE), a phenomenon we identify in the dynamics of the information propagation of social networks, is characterized by a gradual decline in the credibility of the same person who repeatedly spreads and shares information. 
With regard to new participants in an event, the repercussions of the CEE are shown in two aspects. First, if these individuals receive the information from a source with a massive number of supporters, they may doubt the credibility of the information, which could ultimately lead to potential non-participation in the event. Second, if they come to believe in the event for various reasons and decide to participate in the cascade, their belief and participation may not necessarily be directly linked by the most popular person in this propagation cascade.
This does not imply that the popular person with the most supporters lacks the ability to propagate, but rather suggests a need for a comprehensive influence comparison with other participants in the cascade for determining who may be the direct influencer for a new follower of this event, rather than blindly assuming that the new participants are influenced by the most widespread disseminator in the cascade.
(Also, there are some other instances, such as a fervent fan of the widespread disseminator, who maintains high feature similarity or admirer behavior, may remain unaffected by the CEE under any circumstances.)
The CEE can be conceptualized as the "wear and tear" of information upon multiple propagations, which can be interpreted from a view of informational entropy. As information undergoes sharing and spreading, its primality and credibility may be compromised, thereby introducing uncertainty.

\begin{enumerate}
    \item 
    Increment in Informational Entropy: 
With every iteration of information propagation, the entropy (or the uncertainty) associated with that information might escalate. This increase can be attributed to various elements, such as misconceptions, distortions, or omitted segments of the information. When a specific source user spreads identical information repeatedly, a new individual might speculate that such information is more susceptible to "wear and tear", thus casting doubts on its veracity.
Mathematically expressed as:  
\begin{equation}
\label{eq1}
 S'(t) = S(t) + \Delta S,
\end{equation}
   where $ S(t) $ represents the informational entropy at time $ t $, and $ \Delta S $ is the supplementary entropy introduced due to propagation ($\Delta S$ can also be supported and evidenced by the law of entropy increase).

     \item 
   Decay in Credibility of Information: 
The credibility of the information can be perceived as a quantity inversely proportional to its entropy. As entropy is accentuated, the credibility diminishes. Formulated as:  
\begin{equation}
\label{eq2}
    C'(t) = \frac{1}{1 + S'(t)} ,
\end{equation}
   where $ C(t) $ stands for the credibility of the information at time $ t $.

     \item 
     Interrelation of Infection Rate and Credibility:
As the credibility of the information wanes, its infectivity concomitantly dwindles. This mirrors the decline in the degree of trust people place in the information, subsequently reducing the efficacy of its propagation. Expressed as: 
\begin{equation}
\label{eq3}
   R'(t) = R(t) \times C'(t),
\end{equation}
   where $ R(t) $ denotes the infection rate at time $ t $.
\end{enumerate}

This is a fundamental model, primarily elucidating how information gradually loses its credibility during propagation. It's essential to recognize that numerous other variables might influence this trajectory, such as the content of the information, its novelty, and the repute of the disseminator, among others. Nevertheless, the aforementioned framework furnishes us with a foundation, assisting in discerning the essence of the Credibility Erosion Effect in information propagation.
In summary, under regular conditions, the positive influence effect brought by reaching more people and the negative influence effect brought by the CEE collectively shape a disseminator's overall influence within a single event. This influence typically experiences an initial increase, followed by a decrease, and then converges with fluctuations. This pattern is counterintuitive to the commonly held belief that the more people an individual influences, the greater its influence becomes.

\section*{Methods}
\subsection*{Methodology for Complementary Cumulative Distribution Function Statistics}
We employ the CCDFs to analyze various characteristics of rumor cascades and non-rumor cascades on Twitter and Weibo, including the topology features of propagation cascades, directed graph features from the propagation source, and user features of propagation cascades.  The CCDFs metric is a statistical measure used to provide the probability that a random variable is greater than a certain value. It is particularly useful in the analysis of datasets with heavy-tailed distributions, which often appear in social network research~\cite{CCDF}.
The CCDF is calculated as follows:
\begin{equation}
\label{eq4}
   P(X > x) = 1 - F(x),
\end{equation}
where $F(x)$ is the CDF of the feature $X$ under consideration for statistics. 
In our analysis, we first sort the calculated result of the statistical feature $X$ in ascending order. We then calculated the proportion of data points with a value smaller than or equal to each data point in the sorted list $F(x)$, resulting in the CDF. Finally, we can obtain the CCDF based on Eq.~\ref{eq4}.

\subsection*{Methodology for Fitting Cascade Topological Data}~\label{M2}
To gain a deeper understanding of the cascade data, we calculate the topological characteristics for each cascade. Once we have a series of values for each characteristic $\mathcal{F}_j$ from $\mathcal{F} = (\mathcal{F}_1, \mathcal{F}_2, ..., \mathcal{F}_m)$ considering all cascade topologies $G = (G_1, G_2, ... , G_n)$, denoted as $\psi(\mathcal{F}_j)$=\{$\psi(G_1, \mathcal{F}_j)$, $\psi(G_2, \mathcal{F}_j)$,...,$\psi(G_n, \mathcal{F}_j)$\}, we attempt to fit each calculated values ordering $\psi(\mathcal{F}_1)$,...,$\psi(\mathcal{F}_m)$ into various probability distributions, including exponential, gamma, log-normal, and the frequently observed (but not right-skewed) normal distribution. Our methodology revolves around the following approach: for the values ordering $\psi(\mathcal{F}_j)$ associated with a topological feature $\mathcal{F}_j$, we estimate the distribution parameters by employing Maximum Likelihood Estimation (MLE). Here the goal of MLE is to identify a set of parameters that maximize the joint probability of $\psi(\mathcal{F}_j)$ across all cascades.
For a probability density function $f$ parameterized by $\theta_k$ in the estimated distribution $k$, and considering independent and identically distributed (i.i.d.) cascade topologies $G$, the joint likelihood function of feature $\mathcal{F}_j$ is defined as:

\begin{equation}
\label{eq5}
   L_{\mathcal{F}_j}(\theta_k;G,\mathcal{F}_j) = \prod_{i=1}^{n} f(\psi(G_i, \mathcal{F}_j); \theta_k).
\end{equation}

The goal of MLE is to find the parameter $\theta_k$ that maximizes this likelihood function. However, to avoid numerical underflow resulting from the product operation, we typically work with the log-likelihood function. Hence, our objective becomes minimizing the Negative Log-Likelihood Function (NLLF)~\cite{NLLF}:

\begin{equation}
\label{eq6}
   NLLF_{\mathcal{F}_j}(\theta_k;G,\mathcal{F}_j) = -\sum_{i=1}^{n} \log(f(\psi(G_i, \mathcal{F}_j);\theta_k)).
\end{equation}

For each distribution, we identify the parameter estimates that minimized the NLLF, and subsequently computed the NLLF for each distribution with these parameters. Ultimately, we rank all distributions based on the NLLF values, in ascending order. After applying the MLE and minimizing all topological characteristics of NLLF ($\sum_{j=1}^{m} NLLF_{\mathcal{F}_j}$), we conclude that the distribution $k$ that best fits our data is the exponential distribution.

To further assess the extent to which the exponential distribution fits our data, we employ the skewness-based Kolmogorov-Smirnov (K-S) test~\cite{KS}. This test allows us to evaluate the level of agreement between our observed data and the theoretical exponential distribution we have identified as the best fit.
Based on our statistical data $\psi(\mathcal{F}_j)$ related feature $\mathcal{F}_j$, we aim to estimate the parameters of the exponential distribution, namely the location parameter (which accounts for any shift along the x-axis) and the scale parameter (which determines the spread of the distribution). We obtain estimates of the location and scale parameters by maximizing the likelihood function.
\begin{equation}
\label{eq7}
   (\hat{location}, \hat{scale}) = \arg\max_{location, scale} L(location, scale; \psi(\mathcal{F}_j)),
\end{equation}
where $L(location, scale; G)$ represents the likelihood function of $G$ under the exponential distribution parameters, location and scale.
Having obtained the maximum likelihood estimates, we proceed with the Kolmogorov-Smirnov (K-S) test. The K-S test statistic $D$ is defined as:
\begin{equation}
\label{eq8}
   D = \max_{i} |\mathcal{F}_j^{G_i} - \mathcal{F}_j^{E(i; \hat{location}, \hat{scale})}|,
\end{equation}
where $\mathcal{F}_j^{G_i}$ denotes the empirical cumulative distribution function of our statistical data $\psi(\mathcal{F}_j)$ based on each cascade $G_i$ and a given topological characteristic $\mathcal{F}_j$, and $\mathcal{F}_j^{E(i; \hat{location}, \hat{scale})}$ denotes the theoretical cumulative distribution function of the exponential distribution.
We then calculate the associated p-value with the test statistic $D$:
\begin{equation}
\label{eq9}
   p\text{-value} = P(D > d | H_0),
\end{equation}
where $H_0$ represents the null hypothesis that the cascade data follows the exponential distribution.
Certain characteristics of our data, such as structural virality, resulted in p-values exceeding the significance level of 0.05~\cite{anderson2000null}. Consequently, we accepted the null hypothesis, affirming that the exponential distribution is an appropriate fit for these structural characteristics of cascades.

\subsection*{Methodology for User Feature Importance Ranking}
The five categories of user profiles are collected into $\mathbf{X} \in \mathbb{R}^{n \times 5}$, where $n$ represents the total number of cascades. Furthermore, we construct a label vector $\mathbf{y}$ of length $n$, where each entry specifies whether the corresponding cascade is rumor-associated or non-rumor-associated.
To ensure that the chi-square test is not affected by different scales of the features, we first normalize $\mathbf{X}$ to $\mathbf{X'}$ using Min-Max normalization, which transforms each feature to lie within the [0,1] interval~\cite{Min-Max}. This ensures that no feature dominates other features due to its inappropriate scale and each normalization value is non-negative, enabling a fair comparison of the importance of different features in the subsequent chi-squared test.
Subsequently, we perform the chi-squared test on $\mathbf{X'}$ and $\mathbf{y}$ to compute the chi-squared statistic for each feature. This test hypothesizes that each feature is independent of the cascade class, and a smaller p-value suggests stronger evidence against this hypothesis, indicating that the feature is important for distinguishing between the two classes of cascades.
The chi-squared statistic for the $j$-th feature is defined as~\cite{chi-squared}:
\begin{equation}
\label{eq10}
   \chi^2_j = \sum_{i=1}^{v} \left( \frac{(x'_{ij}-\mu_{j0})^2}{\mu_{j0}} + \frac{(x'_{ij}-\mu_{j1})^2}{\mu_{j1}} \right),
\end{equation}
where $x'_{ij}$ is the $i$-th entry cascade of the $j$-th feature after normalization, and $\mu_{j0}$ and $\mu_{j1}$ are the expected values of the $j$-th feature for non-rumor and rumor-associated cascades, respectively.

\subsection*{Methodology for Credibility Erosion Effect Phenomenon}
The discovery of the CEE phenomenon originated from the challenges associated with propagation graph generation. Thus, we would like to elucidate the implied CEE phenomenon in social network propagation by discussing the underlying generative principles and processes. More specifically, 
the scarcity and limited scale of existing propagation data impose significant constraints on the comprehensive evaluation and testing of downstream tasks~\cite{you2018graphrnn,DVAE2019,Nips_SOTA2022_micmac,SOTA2023}, such as influence assessment~\cite{WSDM_deepIS}, source localization~\cite{wang2023lightweight}, user profiling~\cite{AAAI2023}, fake news detection~\cite{yin2023integrating}, and information diffusion analysis~\cite{kumar2021information}. 
To address this challenge, we propose an advanced graph generative model designed to adaptively produce an expanded quantity and larger scale of propagation data. The generated data is closely consistent with the various characteristics and distribution patterns found in real-world propagation data of social media.
The core essence of our generative approach can be decomposed into two intertwined processes: the graph-level and edge-level generations. The graph-level process focuses on characterizing the topology of the current directed acyclic graph. After each update of the latest topological information from the graph-level, the edge-level process starts to measure how to add new participants.  Here, leveraging some techniques (such as the attention mechanism and the recurrent neural network), user attributes are considered to assess the probability of edge formation (i.e., relationship) between a nascent user and each of those already present within the prevailing graph context. The edge structures subsequently devised are reincorporated into the graph-level process, ensuring a cyclical and progressive generation. Such an approach guarantees a structured evolution leading to consistent and coherent graph and edge expansions.
The graph-level generative function $f_G$
and edge-level generative function $f_E$ are defined as follows.
\begin{equation}
\label{eq11}
h_i = f_{G}(h_{i-1}, \Omega(\varphi_{i-1})),
\end{equation}

\begin{equation}
\label{eq12}
\varphi_i = f_{E}(h_i, F),
\end{equation}
where $F$ is the user attributes in the cascade, $h_i$ represents a vector encoding the state of the graph  topology generated so far,  
$\varphi_{i-1}$ is the predicted adjacency vector associated with the most recently generated user $v_{i-1}$, and $\varphi_{i-1}(j)$ ($j<i-1$) signifies the probability of an edge existing between the most recently generated user $v_{i-1}$ and the historical user $v_j$.
$\Omega(x)$ is a one-hot decoder function that sets the maximum value at 1 and sets all other values at 0 from the vector $x$, indicating which historical user is more likely to form an edge with the newly introduced user under the current topological context.
It's worth noting that the design rules for generation ensure flexibility and extensibility. More specifically, a variety of techniques, such as RNN~\cite{GRU}, VAE~\cite{VAE}, GAN~\cite{GAN}, and GCN~\cite{2016GCNICLR}, can be applied to focus on the graph-level generation process $f_G$. And $f_E$ can leverage temporal attention or user attribute attention mechanisms for the generation of new edges. Based on this, the Maximum Mean Discrepancy (MMD) metric, which assesses distributional differences through divergence scores, can validate the distribution discrepancies between propagation data generated from various modules and real-world data~\cite{ICLR_MMD}. A lower MMD score indicates that the generated data closely resembles the characteristics and distribution of real-world data.

\begin{table}[htb]     
    \renewcommand{\arraystretch}{1.0}
    \caption{The generation performance evaluation of different deep methods with or without CEE mechanism based on MMD metric.}
    \label{tab2}
    \centering
    \setlength{\tabcolsep}{3mm}{
    \begin{tabular}{ccccc}
        \toprule    
        Groups& \multicolumn{2}{c}{$\operatorname{CEE~Phenomenon}$} & \multicolumn{2}{c}{$\operatorname{Without~CEE~Phenomenon}$} \\
        \cmidrule(lr){1-1} \cmidrule(lr){2-3} \cmidrule(lr){4-5}
        \textit{Datasets} & \textit{Twitter} & \textit{Weibo} & \textit{Twitter} & \textit{Weibo} \\
        \midrule
        RNN+RNN   & 0.317 & 0.371 & 0.354 & 0.416 \\  
        GCN+RNN & 0.302 & 0.318 & 0.347 & 0.360 \\  
        RNN+VAE & 0.328 & 0.368 & 0.362 & 0.406 \\  
        RNN+GAT & 0.266 & 0.297 & 0.301 & 0.327 \\  
        VAE+GAT & 0.149 & 0.201 & 0.187 & 0.243 \\  
        \bottomrule  
    \end{tabular}}
\end{table}

Inspired by other domains, we have witnessed similar phenomena in areas like advertising~\cite{CEE_instant1} and healthcare~\cite{CEE_instant2}. Therefore, we consider whether such an effect is also prevalent in the propagation of social media.
We attempt to use some tricks to model the CEE phenomenon, 
we introduce a decay mechanism, denoted as $\Psi$, to optimize the edge generation process $\Omega(\Psi(\varphi_{i}))$. After predicting the edge probabilities between a new user $v_j$ and each historical user through $f_E$, these probabilities are adjusted. If a historical user $v_i$ has $k$ succeeding users, then the probability of an edge from $v_i$ to $v_j$ is reduced by a cumulative decay  factor of $\beta^{k}$, where $\beta$ is a decay  factor  very close to but less than 1. 
Subsequently, we carry out rigorous experiments to validate the effectiveness of the decay mechanism, focusing on two primary aspects. Firstly, we employ the MMD metric to compare the discrepancies between generated propagation data with or without the decay mechanism and the actual propagation data. This helped us ascertain whether the CEE phenomenon could make the generated data more analogous to the real-world propagation data. 
Secondly, we compare the performance of downstream tasks using generated data with the decay mechanism against generated data without the decay mechanism.
It is worth mentioning that we do not care which generative model has a smaller MMD or which downstream model performs better. Instead, we are particularly interested in determining whether the presence of the CEE phenomenon indeed enhances experimental performance. Through these rigorous experiments, the existence of the CEE phenomenon can be further validated in the propagation of social media.
Next, we delve into the specifics of these two categories of experiments.
In the first category, as illustrated in Tab.~\ref{tab2}, we adopt several standard techniques from the deep generation domain for our experiments. Remarkably, configurations that incorporated the CEE phenomenon consistently achieve lower MMD scores when comparing within the same datasets or generation methodologies. 
This accentuates the CEE phenomenon's capability in rendering the generated propagation data more analogous to real-world data, indirectly evidencing the presence of the CEE phenomenon in the social network propagation process.
In the second category, we use source localization as an example of numerous downstream tasks of the propagation to explore if considering the CEE phenomenon can enhance model predictive ability in real scenarios~\cite{jiang:2016identifying}. Here, two localization models, GCNSI~\cite{dong:cikm} and TGASI~\cite{ijcai2023}, are used. 
In the experimental setup, the baseline groups were trained on 9/10 of the Twitter propagation data and tested on the remaining 1/10. To set a benchmark, the control group was further enhanced by generating an additional 1,000 real-world propagation graphs without incorporating the CEE phenomenon for training. In contrast, the augmentation group generated another 1,000 real-world propagation graphs, but with the consideration of the CEE phenomenon in the training process. From Tab.~\ref{tab3}, it is evident that the augmentation group, which considers the CEE phenomenon, achieves a higher localization accuracy on the real-world propagation data. 
This suggests that propagation data implying CEE offers a superior capacity to aid models in comprehending the underlying rules of propagation, further emphasizing the importance and presence of the CEE phenomenon in the propagation of social media.

\begin{table}[htb] 
	\renewcommand{\arraystretch}{1.1}
	\caption{Source detection accuracy of localization methods on Twitter under different groups of training sets.}
	\label{tab3}
	\centering
	\setlength{\tabcolsep}{1.5mm}{
	\begin{tabular}{cccc}
		\toprule	
		\textbf{Strategy} & \textbf{Original} & \textbf{Augmentation with CEE} & \textbf{Control without CEE}  \\
		\midrule
            GCNSI~\cite{dong:cikm} & 0.532  & \textbf{0.613}  &  0.566   \\
            TGASI~\cite{ijcai2023}  & 0.787 &\textbf{0.825}  &0.795   \\

		\bottomrule
	\end{tabular}}
\end{table}

\begin{figure}[h]
\centering
\includegraphics[width=0.85\linewidth]{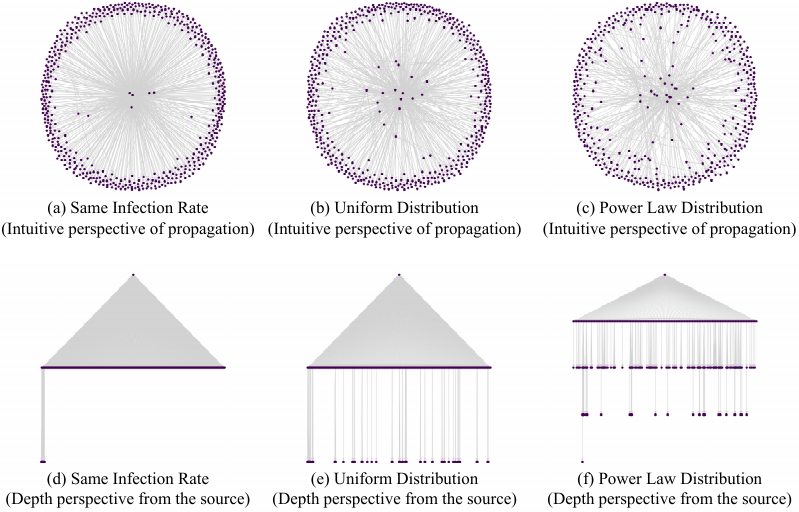}
\caption{Propagation dynamics across three scenarios without considering the CEE phenomenon. 
From left to right, the plots represent:
(1) a homogeneous propagation scenario, where every user has a consistent infection rate,
(2) a heterogeneous scenario with infection rates drawn uniformly at random, and
(3) the Barabási-Albert scenario where infection rates adhere to a power-law distribution.
In all cases, these scenarios are depicted firstly from an intuitive perspective of propagation (the first row) and then from a depth perspective from the source (the second row).}
\label{fig7}
\end{figure}

The above evidence provides an indirect indication of the presence of the CEE phenomenon within social propagation from the perspective of generation tasks.
To rigorously confirm the presence of the CEE phenomenon in social propagation, further discussion and exploration are warranted.
We conducted a comprehensive validation in view of the general propagation dynamics~\cite{chen2022information,yu2021modeling,urena2019review}.
The selected experimental setup obeys the common setting in the mean-filed domain~\cite{mean1999,mean2007,mean_nature}.
Three renowned and authoritative propagation scenarios are introduced to characterize the dynamical patterns from different environments~\cite{wang2020locating}.
\begin{itemize}
    \item Homogeneous propagation scenario: This scenario illustrates a uniform propagation environment where every user has an identical infection rate. Specifically, each user is given an infection rate of 0.5.

    \item Heterogeneous propagation scenario: Here, we account for diverse propagation patterns by assigning each user's infection rate drawn uniformly at random from the range (0,1). This generates a propagation scenario where the infection rate is not fixed but varies randomly among users.

    \item Barabási-Albert propagation scenario: Inspired by the Barabási-Albert model, we emulate a scenario where the infection rate of each user falls within the range of (0,1), adhering to a power-law distribution. This is designed to represent a characteristic of real-world networks in which a few nodes (or users) are highly influential while most others exert minimal influence.
\end{itemize}

\begin{figure}[ht]
\centering
\includegraphics[width=0.85\linewidth]{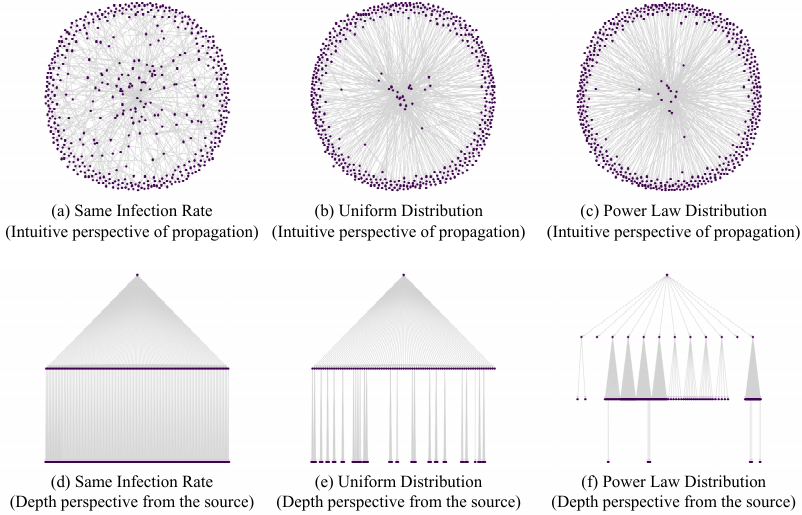}
\caption{Propagation dynamics across three scenarios considering the CEE phenomenon. Other settings are similar to Fig.~\ref{fig7}.}
\label{fig8}
\end{figure}

\begin{table}[htb]     
    \renewcommand{\arraystretch}{1.0}
    \caption{The generation performance evaluation of mean-field propagation dynamics based on different patterns and scales with or without CEE mechanism based on MMD metric.  }
    \label{tab4}
    \centering
    \setlength{\tabcolsep}{1.25mm}{
    \begin{tabular}{cccccccc}
        \toprule    
        Groups& \multicolumn{3}{c}{$\operatorname{CEE~Phenomenon}$} & \multicolumn{3}{c}{$\operatorname{Without~CEE~Phenomenon}$} \\
        \cmidrule(lr){1-1} \cmidrule(lr){2-4} \cmidrule(lr){5-7}
        \textit{Patterns} & \textit{Homogeneous} & \textit{Heterogeneous} & \textit{Barabási-Albert } & \textit{Homogeneous} & \textit{Heterogeneous} & \textit{Barabási-Albert } \\
        \midrule
        100 users   & 0.297  & 0.265 & 0.241 &  0.357 & 0.321 & 0.288 \\  
         500 users & 0.376 & 0.323 & 0.311 & 0.417 & 0.367 & 0.349 \\  
         1000 users& 0.433& 0.392 & 0.379 & 0.474 & 0.423 & 0.420  \\  
        \bottomrule  
    \end{tabular}}
\end{table}

In each propagation scenario, all users initially exist in a stochastic state with no relationship, reflecting a mean-field scenario. 
Any pair of users has the potential to receive a message. Therefore, all users might interact, implying that the initial 10,000 nodes do not have any directed edges. 
Then, as a random source propagates the message, a directed propagation tree ultimately develops.
For each of the three scenarios mentioned above, we ran simulations both by considering the CEE phenomenon and without considering CEE. This led to a total of six distinct groups of propagation trees.
Fig.\ref{fig7} and Fig.\ref{fig8} respectively demonstrate the visualizations of the three scenarios without considering CEE and with considering CEE.
Further, we compute the MMD between each group of the generated propagation trees and real-world propagation data. 
From the results in Tab.~, we observed that for each of the three scenarios, generated propagation data considering the CEE phenomenon consistently has lower MMD scores compared to the corresponding non-CEE variants. This finding underscores that incorporating the CEE phenomenon makes generated propagation data more akin to real propagation data.
From the results in Tab.~, we observed that for each of the three scenarios, generated propagation data considering the CEE phenomenon consistently has lower MMD scores compared to the corresponding non-CEE variants. 
When widely accepted theoretical propagation models, which depict social dynamics, are integrated with the CEE mechanism, there's a distinct resemblance between the generated propagation patterns and the real-world propagation data.
This finding directly underscores that the CEE phenomenon exists in real-world propagation data.


\section*{Data Availability}
We use three datasets collected from two real-world social media platforms, namely Weibo~\cite{weibodata17} and Twitter~\cite{twitterdata1,twitterdata2}
Metadata of the analyzed datasets can be accessed at the following URLs: 
For Weibo, visit \url{https://www.dropbox.com/s/46r50ctrfa0ur1o/rumdect.zip?dl=0} 
and for Twitter, check \url{https://www.dropbox.com/s/7ewzdrbelpmrnxu/rumdetect2017.zip}.
The relevant information of three datasets is shown in Tab.~\ref{datasets}.
Due to platform privacy restrictions, we try our best to publicly submit the statistical analysis data of nearly a million user data across dozens of characteristic dimensions on Twitter and Weibo, as well as the graphic source files (\textit{.opju}) created by \textit{\textbf{Origin}}. 

\begin{table}[htb] 
	\renewcommand{\arraystretch}{1.0}
	\caption{Statistics of the propagation datasets from Weibo and Twitter.}
	\label{datasets}
	\centering
	\setlength{\tabcolsep}{4.5mm}{
	\begin{tabular}{ccc}
		\toprule	
		\textbf{Statistic} & \textbf{Twitter} & \textbf{Weibo}  \\
		\midrule
        \#users    & 770,662 &  2,856,741 \\
        \#cascades & 2,308 &  4664 \\
        \#rumors cascades & 575  &  2244 \\
        \#non-rumors cascades & 1,158 & 2082  \\
        \#user attributes (our work) & 4,623,972 & 17,140,446 \\
		\bottomrule
	\end{tabular}}
\end{table}

The cascade in our study captures two types of propagation: rumors and non-rumors. 
Here, to enhance the accuracy of labeling and reduce the labor intensity of manual annotation, 
the veracity tags of events are found and annotated on the rumor debunking websites (such as snopes.com, Emergent.info, etc)~\cite{weibodata17}.

In both rumors and non-rumors, the directed interaction relationships between users are further captured, such as comments and retweets, forming a successor to a predecessor. To illustrate, consider a scenario where user A is the source who releases a rumor. Users B and C might engage with A's message through comments or retweets. Following this, users D and E could respond to B's message, while users C, D, and E might also receive further engagement through retweets or comments. This series of interactions creates a propagation chain with varying depth and breadth, effectively capturing the dynamic spread of information within the network. What's more, we also analyze the characteristics of all captured cascades, and we find that the user scale ranges from a minimum of 10 to a maximum of 57,186, with an average user scale of 883 across all cascades. Additionally, the cascades vary in propagation depth, ranging from a minimum of 2 to a maximum of 22, with an average depth of 6.15.

\section*{Code Availability}
The codes primarily contain the following several modules:
(1) CCDFs analysis of various dimensions of propagation data;
(2) Exponential distribution fitting of propagation characteristics based on the Kolmogorov-Smirnov test;
(3) User feature importance ranking based on the chi-squared test;
(4) Propagation data generation of social media based on deep graph generative model;
(5) Propagation dynamics diffusion model based on the mean-field considering the CEE.






\end{document}